# Developing a Compton Polarimeter to Measure Polarization of Hard X-Rays in the 50-300 keV Energy Range

J. S. Legere[a], P. Bloser[a], J. R. Macri[a], M. L. McConnell[a], T. Narita[b], J. M. Ryan[a]

[a]Space Science Center, Univ. of New Hampshire, Durham NH 03824
[b]Department of Physics, College of the Holy Cross, Worcester, MA 01610

**ABSTRACT**

This paper discusses the latest progress in the development of GRAPE (Gamma-Ray Polarimeter Experiment), a hard X-ray Compton Polarimeter. The purpose of GRAPE is to measure the polarization of hard X-rays in the 50-300 keV energy range. We are particularly interested in X-rays that are emitted from solar flares and gamma-ray bursts (GRBs). Accurately measuring the polarization of the emitted radiation from these sources will lead, to a better understating of both the emission mechanisms and source geometries. The GRAPE design consists of an array of plastic scintillators surrounding a central high-Z crystal scintillator. We can monitor individual Compton scatters that occur in the plastics and determine whether the photon is photo absorbed by the high-Z crystal or not. A Compton scattered photon that is immediately photo absorbed by the high-Z crystal constitutes a valid event. These valid events provide us with the interaction locations of each incident photon and ultimately produces a modulation pattern for the Compton scattering of the polarized radiation. Comparing with Monte Carlo simulations of a 100% polarized beam, the level of polarization of the measured beam can then be determined. The complete array is mounted on a flat-panel multi-anode photomultiplier tube (MAPMT) that can measure the deposited energies resulting from the photon interactions. The design of the detector allows for a large field-of-view (>$\pi$ steradian), at the same time offering the ability to be close-packed with multiple modules in order to reduce deadspace. We plan to present in this paper the latest laboratory results obtained from GRAPE using partially polarized radiation sources.

## 1. INTRODUCTION

The purpose of GRAPE is to measure the polarization of hard x-rays in the 50-300 keV energy range. As described here, the GRAPE design is most suitable for studies of either gamma-ray bursts or solar flares, as part of a long-duration balloon platform or as part of a satellite platform. There are four properties of the source emitted radiation that can be measured; energy, intensity, direction, and polarization. Many experiments have been conducted to measure the energy, intensity and direction of these incoming photons, but measurements of polarization have been lacking largely due to the lack of instrumentation with sufficient sensitivity. Polarization measurements have become a powerful tool for astronomers throughout the electromagnetic spectrum. It is believed that by accurately measuring polarization levels from solar flares and GRBs we will be able to better understand both the emission mechanisms and source geometries producing the observed radiation.[1]

Mechanisms responsible for the emission of gamma rays are mainly non-thermal and many of which can lead to high degrees of linear polarization while being dependent on the geometry of the source.[1] Four such mechanisms are 1) Magneto-Bremsstrahlung radiation, including (cyclotron, synchrotron, and curvature radiation); 2) Compton Scattering, which can polarize initially unpolarized photons; 3) electron-proton Bremsstrahlung radiation can produce levels of linear polarization up to 80%; and 4) Magnetic photon splitting can lead to levels up to 30%. Each of these mechanisms will emit gamma rays with a known level of polarization at hard X-ray energies. Measuring the polarization levels from the emissions of solar flares and GRB, among other sources, will help us determine if the particle accelerations are thermal in nature or closely resemble one of the models suggested.

A solar flare occurs when the magnetic energy that is built up in the solar atmosphere is suddenly released. Flares generally occur near sunspots and the two are thought to be closely related. During this energy release, high-energy subatomic particles, mostly electrons, protons and neutrons are accelerated. The ions can be accelerated up to tens of GeV and the electrons up to hundreds of MeV. Along with high-energy particles and optical emissions, solar flares also emit x-rays and gamma rays. The study of polarization of these x-rays and gamma rays emitted from solar flares will help in determining the mechanisms behind the rapid accelerations of ions and electrons that are currently unknown. Polarization measurements of solar flares are expected to be useful in determining the extent to which the accelerated electrons are beamed, which in turn, has important implications for particle acceleration models.[2-10]

Another class of object that will be observed for polarization levels by GRAPE are gamma-ray bursts (GRBs). Being the brightest source of gamma rays known to date, they are now believed to be located in distant galaxies. These great distances infer that GRBs emit energies of $10^{51}$-$10^{53}$ ergs or more over the span of several seconds. The current model of GRBs is one that describes this release in terms of a relativistic fireball model. According to the current fireball model, gamma rays are emitted when an ultra-relativistic energy flow is converted into radiation.[11,12]

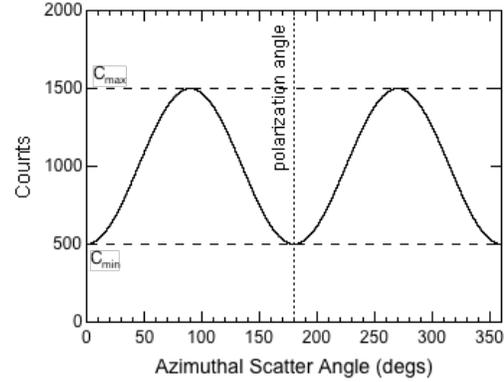

The mechanism of the "inner engine" is of particular interest. There have been numerous models proposed to describe this "inner engine". Most tend to describe the release as a massive short-lived accretion disk surrounding a black hole. Another popular model is one that involves the release of energy along jets that are directed along the rotation axis. Whatever the model it is expected that accurate polarization measurements will give a clearer picture as to the energy release mechanism and the source geometry.

Figure 1: The modulation pattern observed when Compton scattering polarized radiation.

In this paper we plan to present a brief background in Compton polarimetry followed by a brief history of GRAPE and results for previous laboratory science models. We then discuss the current data from the third GRAPE science model. The future of the GRAPE design and possible deployment options are then reviewed.

## 2. COMPTON POLARIMETRY

Measuring the polarization of X-rays in the 50-300 keV energy range is most easily done by Compton scattering. The basic design of a Compton polarimeter is one that consists of two detectors.[1,13] The detectors are capable of determining the energies of the scattered photon and recoil electron along with the location of the Compton interaction site and direction of the scattered photon. The first, a low-Z (scattering) detector, has a high probability of Compton interactions, and the second, a high-Z detector (calorimeter) that has a high probability of absorbing the scattered photon. By analyzing the kinematics of each event we can determine the path of each photon that interacts within the detector. For an incident photon with frequency $v_0$ the Compton scattering differential cross section is given by the Klein-Nishina formula,

$$d\sigma = \frac{r_o^2}{2} d\Omega \left(\frac{v'}{v_o}\right)^2 \left(\frac{v_o}{v'} + \frac{v'}{v_o} - 2\sin^2\theta \cos^2\eta\right) \quad (1)$$

where,

$$\frac{v'}{v_o} = \frac{1}{1+\left(\frac{hv_o}{mc^2}\right)(1-\cos\theta)} \quad (2)$$

with ν′ the frequency of the scattered photon. The angle θ is the scattering angle of the scattered photon measured from the direction of the incident photon, and η is the azimuthal angle of the scattered photon measured from the plane containing the electric field vector of the incident photon.

Investigating Equation 1 we can see that for a fixed value of θ the scattering cross section for polarized photons is a maximum at η = 90° and a minimum at η = 0°. This leads to an asymmetry in the number of photons scattered in the direction parallel and orthogonal to the incident photon electric vector. In other words, photons will tend to scatter 90° to their incident electric field vector. It is this phenomenon that is utilized in the Compton polarimeter.

The polarization measurement is accomplished by recording the azimuthal modulation pattern of the scattered photons. We bin each valid scattering event by its corresponding azimuthal angle. This produces a modulation pattern that can be fit with,

$$C(\eta) = A \cos\left(2\left(\eta - \varphi + \frac{\pi}{2}\right)\right) + B \qquad (3)$$

where φ is the polarization angle of the incident photons and A and B are constants used in the fit. With a proper modulation fit we can use the parameters A and B to find the polarization modulation factor $\mu_p$.

$$\mu_p = \frac{C_{p,\max} - C_{p,\min}}{C_{p,\max} + C_{p,\min}} = \frac{A}{B} \qquad (4)$$

$C_{p,\max}$ and $C_{p,\min}$ represent the maximum and minimum number of counts measured in the polarimeter, Fig 1. The quality of the polarization signature is quantified by the polarization modulation factor.[1,14]

In order to determine the polarization of the incident beam we need to know the response of the detector to a beam of 100% polarized photons. This can be measured experimentally or by performing Monte Carlo simulations for the particular detector design and structure. The modulation factor for a completely polarized beam ($\mu_{100}$) can be used to find the unknown polarization P,

$$P = \frac{\mu_p}{\mu_{100}} \qquad (5)$$

To determine the minimum detectable polarization (MDP) of a specific detector design we use,

$$MDP(\%) = \frac{n_\sigma}{\mu_{100} R_{src}} \sqrt{\frac{2(R_{src} + R_{bgd})}{T}} \qquad (6)$$

where $n_\sigma$ is the significance level (number of sigma), $R_{src}$ is the total source counting rate, $R_{bgd}$ is the total background counting rate and T is the observation time. In terms of detector development, increasing the modulation factor for 100% polarized measurement $\mu_{100}$ or the source count rate $R_{src}$ (increasing the effective area of the detector), will decrease the MDP of the detector.

# 3. INITIAL CONCEPT

The design of a Compton polarimeter is subject to specific criteria. The ability of the detector to reconstruct the kinematics of single scattering events, while eliminating multiple scattering events, is of ultimate importance. As mentioned earlier, GRAPE utilizes a design that incorporates two detectors; an array of scattering detectors surrounding a central calorimeter detector. In event reconstruction the measured parameters include the location of the Compton interaction, the energy of the scattered electron and the energy of the scattered photon. The scattering detectors provide us with the Compton scatter location (in reference to the central calorimeter) of the incident photon and the energy of the recoil electron. The scattered photon is then fully absorbed by the calorimeter and its energy recorded. The ability of the detector to determine the modulation pattern, and hence the polarization parameters, is directly determined by the accuracy with which we can determine this kinematic information.

## 3.1 Principal of Operation

The development and design of the GRAPE detector has evolved through three science models.[2-10] Each one represented a successive improvement, but all three essentially operate under the same underlying principle of operation. A high-Z material, the calorimeter, is surrounded by multiple plastic scintillation detectors that serve as a target for the Compton scattering. The entire array is contained within a single light-tight housing. The plastic scintillators are made of a low-Z material that maximizes the probability of a Compton interaction. The purpose of the calorimeter is to fully absorb the energy of the scattered photon. Ideally, photons that are incident on the plastic scintillator array will Compton scatter only once, and then be subsequently absorbed by the calorimeter. For such an event we measure the energy of the scattered electron in the plastic and the deposited energy of the scattered photon in the calorimeter. With multiple plastic scintillators surrounding the calorimeter, we can determine the azimuthal scatter angle of each valid event. A histogram of these data represents the azimuthal modulation pattern of the scattered photons, which provides a measurement of the polarization parameters (magnitude of the polarization and polarization angle) of the incident flux.

In order to accurately measure the azimuthal modulation (and hence the polarization parameters), we need to correct for geometric effects specific to the individual detector design. When the azimuthal modulation profile is generated, the distribution not only includes the intrinsic modulation pattern due to the Compton scattering process, but it also includes various geometric effects. One of these effects originates from the specific layout of the detector elements within the polarimeter and the associated quantization of possible scatter angles. Other effects include such things as the nonuniform detection efficiency of the PMT used for detector readout. In order to correctly analyze the data we first measure unpolarized radiation with the polarimeter. These data provide a measure of the various geometric effects. By dividing the polarized distribution by the unpolarized distribution and then normalizing with the average of the unpolarized distribution we obtain a corrected distribution with the geometric effects eliminated. In the laboratory, these effects are easy to measure using data from an unpolarized beam. In practice, simulations may be used to determine the unpolarized modulation pattern that is used to correct the measured data.

To determine the polarization level, P, of the incident radiation we need to know the modulation factor for a completely polarized beam ($\mu_{100}$). We have used simulations based on MGEANT (incorporating the GLEPS polarization code) to model the response of the polarimeter to 100% polarized incident radiation. The simulations included all important components of the lab setup.

## 3.2 Science Model 1 (PSPMT design)

Science Model 1 (SM1) was designed so that the entire device fit on the front end of a 5-inch diameter position-sensitive PMT (PSPMT) (Hamamatsu R3292).[2-4] The scattering elements were 250 plastic scintillators, Bicron BC-404, each 5mm × 5mm × 50mm. Each was wrapped in Tyvek® in order to maximize the light collecting ability of each plastic and to eliminate cross talk between elements. Kapton® tape was used to secure the Tyvek® wrap on each plastic. The array of plastic elements was arranged, Fig. 3, with a square void in the center allowing for the insertion of the calorimeter. The calorimeter was a 2 × 2 array of 1cm × 1cm × 5cm CsI(Na) elements coupled to a 2 × 2 multi-anode PMT (MAPMT) (Hamamatsu R5900). The calorimeter detector was enclosed in its own light-tight housing and placed within the central void of the plastic array.

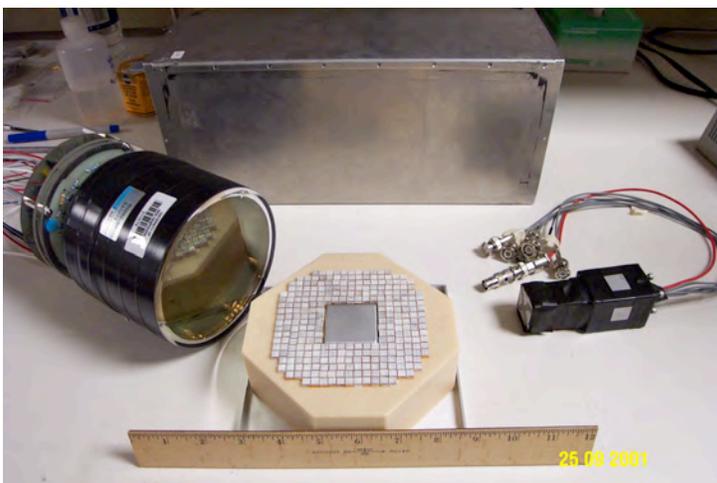

Figure 2: The SM1 detector with the PSPMT (left), the plastic scintillator array (middle), and the calorimeter assembly (right).

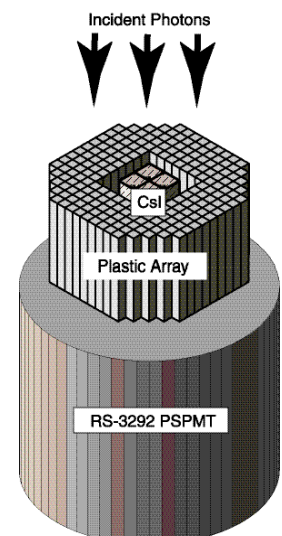

Figure 3: The SM1 design concept.

The PSPMT made use of a charge division network for readout that provides a weighted average of the spatial distribution of the measured light output using only four signals (two in x and two in the y). The MAPMT also has four outputs, one for each of the anodes coupled to a single CsI(Na) crystal. Valid events are recorded for a single Compton scatter occurring in a plastic and a single photo-absorption occurring in a CsI(Na) crystal.

Experiments were conducted in the laboratory in order to determine the effectiveness of the detector. The polarized radiation used in the tests was produced by Compton scattering radiation from a laboratory gamma-ray source.[2,16] Radiation from a $^{137}$Cs source is collimated with lead shielding and directed toward a plastic scintillation detector, referred to as the polarizer. Photons with an initial energy of 662 keV scatter through an angle of 90° before reaching the polarimeter detector. The radiation reaching the polarimeter is highly polarized (at a level of ~55-60%) and has a reduced energy of 288 keV. The use of the polarizer also allows us to electronically tag photons. A triple coincidence between the polarizer and the two sets of polarimeter detectors provides an efficient means for recording data from the polarized beam. The level of polarization measured by SM1 was 55(±2)%, in agreement with the expected value of 55-60%. These results verified both the operation of the SM1 prototype and the efficacy our computer modeling.[7]

## 4. AN IMPROVED DESIGN

### 4.1 MAPMT Design

Following on the success of SM1 we developed a more compact design based on the use of flat-panel MAPMT (Hamamatsu H8500).[9] The H8500 is a MAPMT with an array of 8 × 8 independent anodes. The 5 mm anodes are arranged with a pitch of 6 mm, occupying a total area of 52 mm × 52 mm. With a depth of 28 mm, the flat-panel PMT is far more compact than the PAPMT used in SM1.

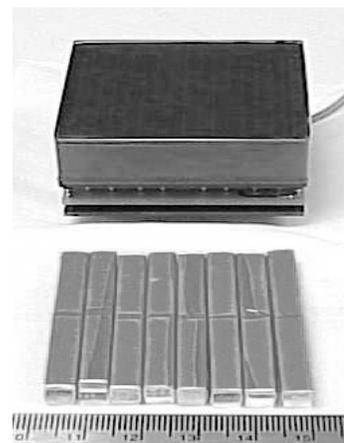

Figure 4: The Hamamatsu H8500 MAPMT (top), and several plastic scintillator elements (bottom).

The MAPMT design offers several advantages over the SM1 design.[9] One advantage is that the detector can be contained in a single housing with one readout device. There is no need to use a second detector for the calorimeter; the complete array is coupled directly to the face of the H8500. This eliminates the need for second HV line and reduces the overall mass of the polarimeter. More importantly it eliminates the obstructive mass of the MAPMT used for the readout of the calorimeter array in the SM1 design. The square design of the MAPMT also allows for the close packing of multiple modules with minimal dead space. Compared to the SM1, an array of four MAPMT modules provides approximately 60% larger effective area and a 30% smaller footprint. The much smaller depth of the MAPMT design also results in a smaller volume.

### 4.1.1 Science Model 2 (SM2)

First tests of the flat-panel MAPMT design, science model 2 (SM2), were done with 12 plastic scintillators measuring 1 cm × 1cm × 5 cm, all wrapped in Tyvek® and secured with Kapton® tape. Each plastic scintillator was read out with a group of four MAPMT anodes. The signals from within each group of four anodes were summed together, providing a total of 12 data channels for the scattering elements, Fig. 6. The center four anodes were used to read out the CsI(NaI) calorimeter. The whole array assembly was held together with cookie spacers made from Delrin® and coupled to the face of the MAPMT with optical grease. This setup allowed for initial measurements with the new MAPMT design to evaluate some of the fundamental characteristics of the design without significant changes to the data acquisition setup.

Laboratory tests of SM2 were conducted during the summer of 2004 with polarized radiation at two different polarization angles, 0° and 90°. The polarized radiation was created in the lab according to Sec. 3.2 and the data was corrected as described in Sec. 3.1. For these measurements the polarized beam was vertically incident on the front surface of the polarimeter. The corrected plots for the two polarization angles are shown in Fig. 5.

For the data collected at 0° polarization angle, a fit with Equation 3 yielded a modulation factor of 0.28(±)0.06, corresponding to a measured polarization level of 55(±14)%, in agreement with the 55-60% value expected. For the data collected at a 90° polarization angle, the measured polarization level was 53(±9)%. Simulations done with MGEANT predicted $\mu_{100}$ for this detector configuration to be 0.48(±)0.03, which was used to derive these polarization levels.

### 4.1.2 Science Model 3 (SM3)

With the first tests completed of SM2 successfully, the MAPMT was fully populated with 60 plastic scintillators according to the original MAPMT design. As with SM1, the plastic scintillators are wrapped in Tyvek® and secured with Kapton® tape. In order to ensure proper spacing of the elements on the MAPMT anodes, cookies made of Delrin® were fabricated to hold the array together as one complete piece, Fig. 7. The whole array is coupled to the MAPMT using clear G&E silicon. This was done after testing revealed the amount of air bubbles present when optical grease was used. Using the silicon greatly reduced the number of air bubbles and in turn increased the light output measured in the MAPMT anodes. The silicon also offers the advantage of being more secure, making the handling of the detector easier in the lab. The whole detector is housed within a light tight aluminum and Delrin® case seen in Fig. 6.

Initial laboratory tests of SM3 proved unsuccessful. It was discovered that there were light cross-talk issues between the calorimeter and the twelve surrounding plastic elements. We believe that this can be attributed to the close proximity between the calorimeter and adjacent plastics. With a light output of ~4 times that of a plastic scintillator, the CsI(NaI) calorimeter was contributing to the energy recorded in this inner plastic ring. This virtually eliminates our ability to reconstruct the kinematics of valid events.

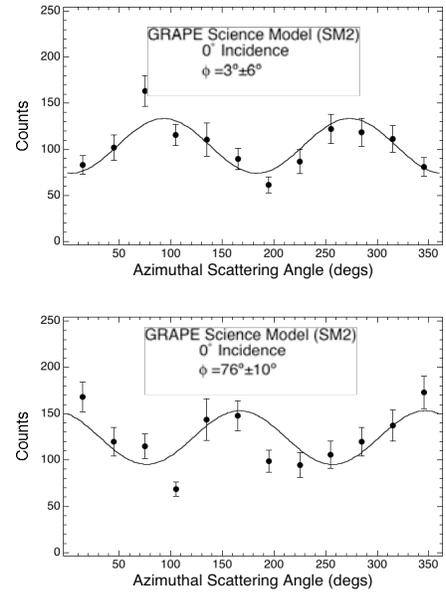

Figure 5: Results from SM2 showing the 90° shift in the modulation pattern when rotating the detector 90° to the incident polarization vector.

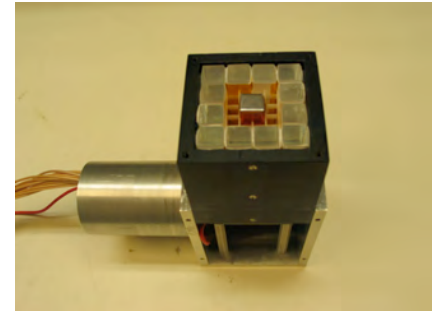

Figure 6: SM2 in initial testing phase with large plastics populating four anodes for a total of 12 scattering elements.

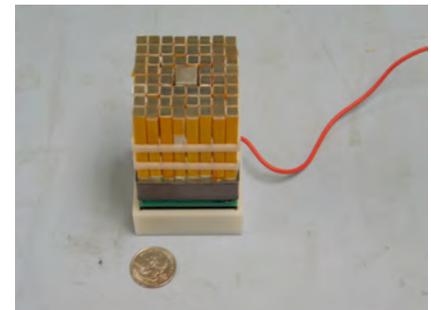

Figure 7: SM3 fully populated with 60 plastics and central CsI(NaI) calorimeter.

To correct for the cross-talk problem we decided to eliminate the inner twelve plastic scintillator elements from the analysis. This leaves 48 plastic elements in the array for analysis of the azimuthal scatter angle. The results reported in Fig. 8 represent the data collected with inner ring in place, but events occurring in the inner ring eliminated. Data was collected for 0° and 90° polarization angle and yielded polarization levels of 56(±9)% and 55(±7)% respectively. Again simulations were used to obtain $\mu_{100}$ allowing us to determine these measured levels.

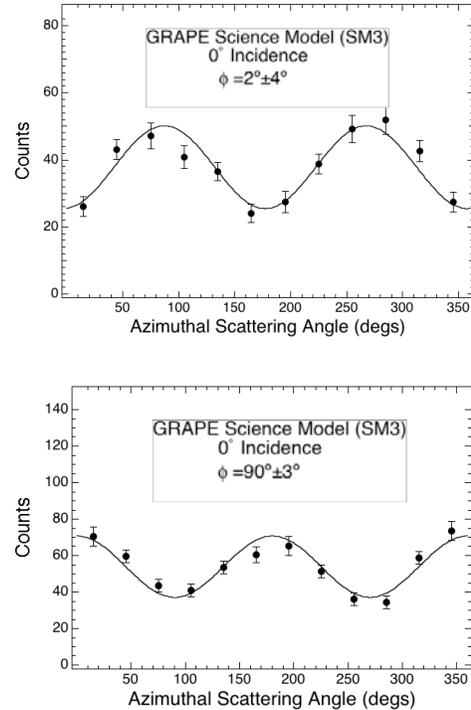

Figure 8: Results from SM3 with inner plastic ring eliminated.

### 4.2 Hardware
Along with the changes in detector design from SM2 to SM3 we have also modified the data acquisition system to handle the increased number of data channels. The number of data channels has increased from 13 (in SM1) to 61 (in SM3).

Preamp boards were designed to be attached to the output pin assembly on the base of the MAPMT, Fig. 9. Each preamp board handles 16 anode channels plus 1 test channel. All signals are processed in CAEN spectroscopy amplifiers (MOD. N568), which splits each signal into fast and slow outputs. The fast outputs are used for timing purposes and the slow outputs are used for pulse height analysis.

Each slow signal from the MAPMT is sent to the ADC's for pulse height conversion. Also housed in the VME crate is a CAEN constant fraction discriminator (CFD, CAEN V812) with 16-channel capability. In order to use the CFD to monitor the 60 fast signals from the plastic scintillators, a passive sum box was designed and built to combine the signals from the plastic scintillators into 15 groups. The gain-matched fast signals from each group were sent to the sum box and the single sum output sent to the CFD. The CFD has an OR output option, which functions by outputting a logic signal when one of its 16-channels is above the programmed threshold. This works because we are interested in single events occurring in any one plastic. When a signal above the CFD threshold occurs in any one plastic group, a logic signal is sent to the logic portion of the setup for coincidence determination.

The VME-based CFD only monitors the fast plastic signals. Separate NIM-based CFDs are used for the polarizer and calorimeter channel. The NIM-based CFDs are adjusted individually for threshold and delays. Each CFD will then send a logic signal to the coincidence portion of the setup when programmable threshold levels are exceeded for each detector.

The coincidence (logic) piece of the setup consists of 1) a PLS 794 Quad gate generator 2) a LeCroy 365AL 4-Fold Logic Unit 3) a LeCroy 222 Dual gate generator. The three logic signals (plastic, CsI(NaI), polarizer) are input to the PLS 794 in order to create gates for each channel. The gate widths are set according to the speed of each detector, the CsI(NaI) having a much slower rise time compared to the plastics. These gates are fed to the logic unit where an AND circuit allows the user to select the level of coincidence. For example, a polarized run would be set to have the three gates active, requiring a triple coincidence level. When energy calibrations are needed, the appropriate detector can be selected and the level of coincidence set to single. The output of the logic unit is sent to the LeCroy gate generator, which generates a gate that is used to trigger ADC conversion.

### 4.3 Software

The CAEN M785 ADCs have the capability to store 32 events (32 pulse height conversions), in their output buffer. Controlling the operation of the VME crate, the two ADC modules and the CFD module is achieved using KMAX version 8.03 from Sparrow Corp. KMAX allows control of the VME crate through the SBS 620 controller VME card and PCI board. Our setup uses a Macintosh G4 computer running MacOSX (v10.3). KMAX communicates to VME modules through a fiber optic cable connected from the SBS PCI card in the Mac to the SBS controller module housed in the VME crate.

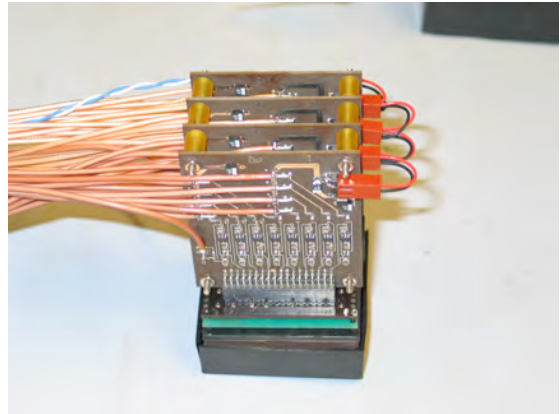

Figure 9: The preamp circuit boards plugged into the back end of the MAPMT.

The ADC's needed to be properly programmed before they can function properly. The proper sequence is: 1) reset ADC; 2) activate all channels; 3) disable zero suppression; 4) setting of all channel thresholds to zero. Disabling zero suppression allows the ADC to report all 32 channels of data. All channel thresholds are set to zero because we need all channel data during an event. The majority of the programming effort lies with acquiring the data from the output buffers of the ADC.

Due to the fact that we are using two ADC's to acquire data, synchronization becomes a very important factor. When the ADC's 32 event output buffer is full a flag is written at a specific address. KMAX polls this address and waits for the flag. When it reads the flag KMAX initiates the acquisition portion of the program. The ADC's are shut off and all events are read from the buffers. The ADC's tag all data with an event number and it is this number that is used to determine the synchronization. If the ADC's are off in event numbers the event block is deleted as unusable data. This can happen when the ADC's are turned on. If the event rate is high enough or just a random coincidence happens, the first ADC can receive a gate before the second one is turned on. This requires that all data be checked for synchronization prior to writing to the data file. After the data is read the ADC's are again turned on to begin acquiring data. During our polarization experimental runs this problem has not been evident, but during energy calibrations it can be a problem if event rates are too high.

Other programming involved with Kmax is the initialization and setup of the VME CFD. Similar to the ADC's the CFD must be reset and all channels activated. Each channel is programmed with a threshold value specific to its gain. When there is a valid event occurring in any plastic group the fast signal exceeds the programmed threshold and a logic signal is sent to the gate generator.

The data written in the event files is analyzed using the ROOT software package. It is with ROOT that we apply the energy calibration to each detector. We can then look at each event and determine its validity. As stated earlier, we are interested in events occurring in single plastics in coincidence with events in the calorimeter and the polarizer. Once a valid event is identified the scattering angle is recorded and histogrammed.

## 5. GRAPE FUTURE

### 5.1 Further SM3 Testing

With the successful testing of SM3 nearing completion, the future of the GRAPE module looks promising. We first plan to test the SM3 with the inner ring of plastic scintillators removed. This should increase detector sensitivity by improving the statistics of valid events measured in polarized tests. We also plan to test GRAPE at various incidence angles of incoming polarized radiation in order to determine sensitivity of the polarimeter to off-axis incident radiation.

## 5.2 Science Model 4 (SM4)

The next step for GRAPE is improved coincidence timing and energy resolution. We plan to accomplish this by replacing the central CsI calorimeter with one based on Lanthanum Bromide (LaBr$_3$).[16] This relatively new inorganic scintillator provides an energy resolution that is more than twice as good as NaI at 662 keV (3% vs. 7%). The LaBr$_3$ also provides for better timing characteristics. With decay times of ~25 nsec, it is comparable to the plastic scintillators used and will greatly improve the coincidence timing.

## 5.3 GRAPE Deployment

In order to provide adequate sensitivity, any realistic application of the GRAPE design would involve an array of polarimeter modules.[9] One possible deployment option for a GRAPE polarimeter array would be as the primary instrument on an Ultra-Long Duration Balloon (ULDB) payload. The ULDB technology currently under development by NASA is expected to provide balloon flight durations of up to ~100 days. A 1 m$^2$ array of GRAPE modules would easily fit within the envelope of a ULDB payload. The ideal configuration for GRB studies would be an array that remains pointed in the vertical direction (i.e., towards the zenith) at all times. In this case, there would be no pointing requirements, only a moderate level of aspect information (continuous knowledge of the azimuthal orientation to ~0.5°). An imaging polarimeter could also be designed to match the payload limitations of a ULDB, although the pointing requirements would be much more severe (<1° in both azimuth and zenith). A second deployment option would be as part of a spacecraft payload. This could be either a free-flying satellite or an add-on experiment for the International Space Station (ISS).

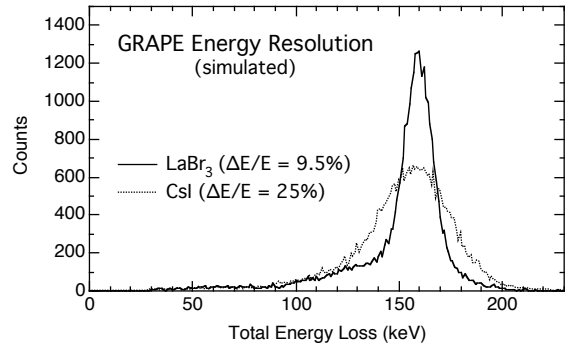

Figure 10: Simulated results showing the improved energy resolution of the SM4 design that would result from changing the CsI calorimeter to one made from LaBr$_3$.

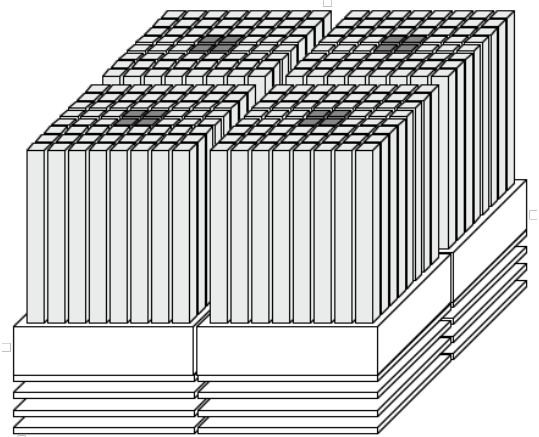

Figure 11: An array of four GRAPE modules.

## ACKNOWLEDGEMENTS

This work is currently supported by NASA grants NNG04GB83G and NNG04WC16G. We would like to thank Mark Widholm and Paul Vachon for their support with the MAPMT design electronics. We would also like to thank the Laboratory for Advanced Instrumentation Research at Embry Riddle, Sparrow Corp., Drew Weisenberger, and Sergio Brambilla for their support with the VME data acquisition setup.